\documentclass[12pt]{iopart}
\expandafter\let\csname equation*\endcsname\relax
\expandafter\let\csname endequation*\endcsname\relax
\usepackage{amsmath}
\usepackage[english]{babel}
\usepackage{bm}
\usepackage{bbm}
\usepackage{iopams}
\usepackage{graphicx}
\usepackage{color}

\usepackage[
  breaklinks=true,
  colorlinks=true,
  linkcolor=blue,
  urlcolor=blue,
  citecolor=blue
]{hyperref}
 \usepackage[numbers,sort&compress]{natbib}
 \usepackage{hypernat}
 
\begin{document}
\title[]{Signatures of relativistic spin-light coupling in ultrafast magneto-optical pump-probe experiments}
\author{Ritwik Mondal, Marco Berritta and  Peter M. Oppeneer}

\address{Department of Physics and Astronomy, Uppsala University, P.\,O.\ Box 516, Uppsala, SE-75120, Sweden}
\ead{ritwik.mondal@physics.uu.se}
\begin{abstract}
Femtosecond magneto-optical pump-probe measurements of ultrafast demagnetization show an intriguing difference in the first 100 fs of the magneto-optical Kerr response 
depending on whether the polarization of the pump and probe beams are in parallel or perpendicular configuration [Bigot {\it et al.}, Nature Phys. {\bf 5}, 515 (2009)]. Starting from a most general relativistic Hamiltonian we focus on the ultra-relativistic light-spin interaction and show that this coupling term leads to different light-induced opto-magnetic fields when pump and probe polarization are parallel and perpendicular to each other, providing thus an explanation for the measurements. We also analyze other pump-probe configurations where the pump laser is circularly polarized and the employed probe contains only linearly polarized light and show that similar opto-magnetic effects can be anticipated.

\end{abstract}
\date{\today}
\section{Introduction}
Magneto-optical measurements  have been instrumental to understand many physical and chemical properties of magnetic materials \cite{Buschow1988,Yan1989,Kato2004,Crooker2005,Schemm2014}.
These phenomena can be measured accurately by means of the Kerr rotation and Kerr ellipticity in magneto-optical Kerr effect (MOKE) measurements in reflection geometry or as the Faraday effect in transmission geometry \cite{Schoenes1992,zvezdin97}.
The MOKE and Faraday effects are relatively small effects that originate from the relativistic spin-orbit coupling \cite{argyres55,oppeneer1992ab}. Notwithstanding the smallness of magneto-optical effects they can be very well described by using \textit{ab initio} electronic structure calculations on the basis of the relativistic density functional theory (see, e.g., \cite{kraft95,oppeneer01}).

In the last two decades magneto-optical spectroscopy has been implemented in pump-probe experiments, employing ultrashort laser pulses in the subpicosecond regime to excite a magnetic material and subsequently monitor the spin dynamics with time-resolved MOKE measurements (TR-MOKE)
\cite{beaurepaire96,koopmans00,Bigot2002,vankampen2002,kirilyuk10}. 
The probing of the magnetization response was initially performed at optical wavelengths (see, e.g.\  \cite{kirilyuk10} for a review) but 
more recently X-ray magneto-optical spectroscopy became employed to detect the magnetization response in an element-selective way in magnetic alloys \cite{stamm07,radu11,Mathias12,rudolf12,Vodungbo12,bergeard14}.
These pump-probe measurements revealed an extremely fast decay of the magnetization of a laser excited ferromagnetic film \cite{beaurepaire96,Bigot2002} much faster than what was expected from spin-lattice relaxation times \cite{Vaterlaus1991,wang10}.
The origin of the surprisingly fast laser-induced spin dynamics has been much debated in recent years \cite{carpene08,krauss09,Koopmans2010,Zhang2009,battiato10,carva11nat,faehnle11,mueller13,Krieger2015}. Among the currently most discussed mechanisms are Elliott-Yafet electron-phonon spin-flip scattering 
\cite{Koopmans2010,Carva11,essert11,carva13}, ultrafast magnon generation \cite{carpene08,haag14,turgut2016}, and spin transport through superdiffusive spin currents \cite{battiato10,battiato12}.
Laser-induced magnetization dynamics on longer time scales---as given by the near thermalization of the involved sub-systems lattice, spins and electrons---can be relatively well described by simulations on the basis of the Landau-Lifshitz-Gilbert equation  combined with a two- or three-temperature model \cite{atxitia07,atxitia10,ostler12,Baryakhtar13,ellis15}. The processes occurring on ultrashort timescales, immediately after and during the pump excitation, are much less understood. Bigot \textit{et al.} performed a TR-MOKE experiment in which they employed different arrangements of the pump and probe linear polarizations \cite{bigot09}. They measured the Kerr rotations for the linearly polarized pump beam, setting its polarization either parallel $\theta_{\rm PP} =0^{\circ}$ or perpendicular $\theta_{\rm PP}= 90^{\circ}$ to the linearly \textit{s}-polarized probe beam, and observed an intriguing difference in the differential Kerr rotations as function of the pump-probe delay $\tau$ on a femtosecond timescale, see Fig.\ 1.
The difference in the demagnetization observed for the two configurations was attributed to a \textit{coherent} contribution to the ultrafast magnetization dynamics \cite{bigot09}, occurring at short delay times of about 100 fs. In the demagnetization process there is additionally an incoherent part that does not depend on the polarization of the pump and probe beams, such as e.g.\ spin-flip scattering or spin transport processes of hot electrons \cite{bigot09,bovensiepen09}.  
Notably, at such short time delay there is a non-negligible overlap of the pump and probe beams, suggesting that the interaction of these beams plays a role. The coherent contribution to the magnetization dynamics  was attributed to relativistic quantum electrodynamics stemming from relativistic effects beyond the common crystalline spin-orbit interaction \cite{bigot09,hinschberger13,Hervieux2016}.
\begin{figure}[h!]
\centering
\includegraphics[width = 12cm]{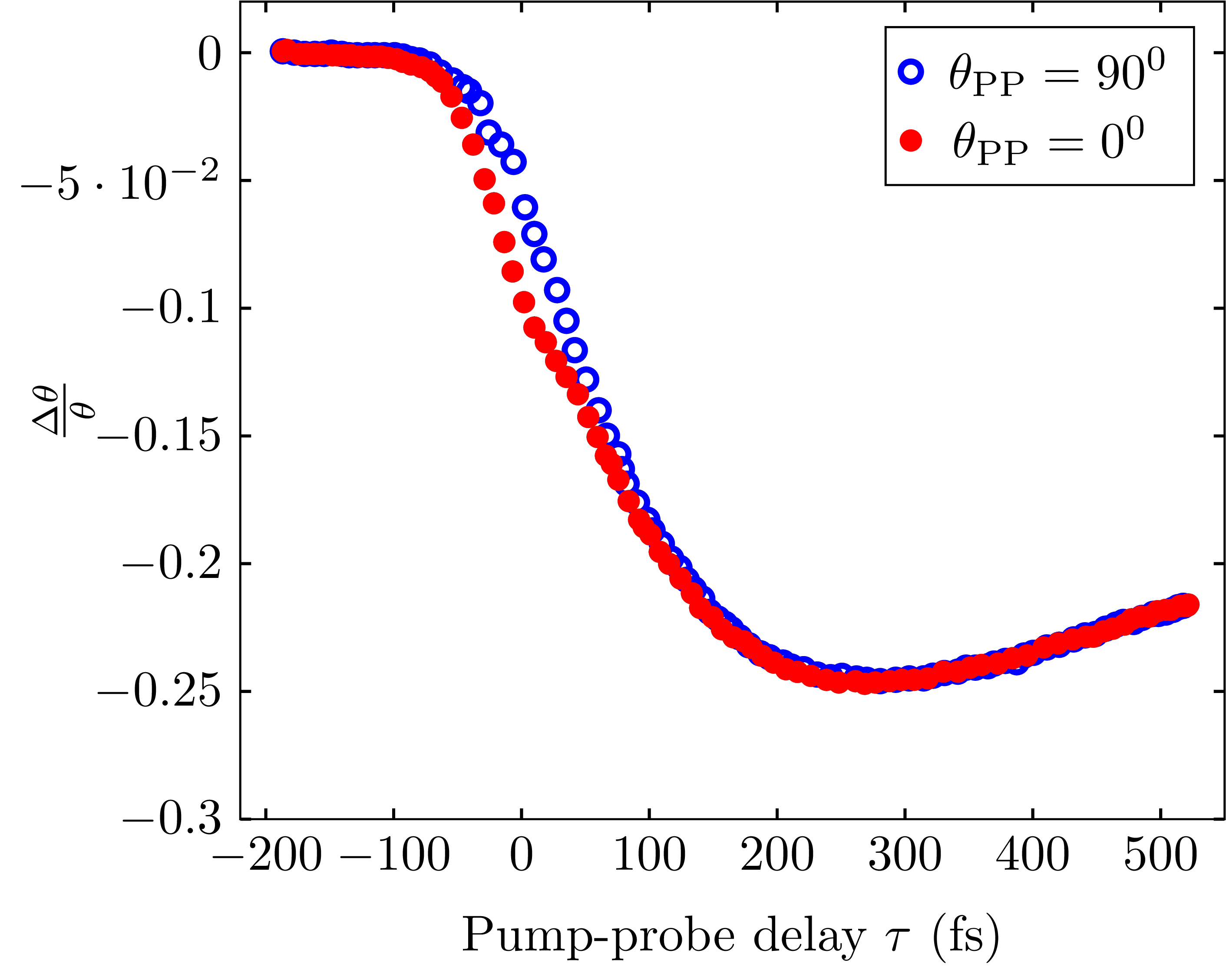}
\caption{(Color online) The normalized differential Kerr angle measured for a thin nickel film, for parallel and perpendicular polarization of the pump and probe beams. The data have been taken from Bigot \textit{et al.}\ \cite{bigot09}.}
\label{fig1}
\end{figure}
It is already well established that the magneto-optical Kerr effect is a relativistic quantum effect because of its relation to the crystalline spin-orbit coupling \cite{argyres55,oppeneer1992abso,huhne99}. In an earlier work we investigated the influence of additional relativistic effects, including the exchange field, laser field and induced relativistic spin-flip optical transitions, and calculated \textit{ab initio} their contribution to the linear MOKE response \cite{Mondal2015a};
we found that these additional relativistic contributions do not lead to any significant change in the MOKE spectra of Ni \cite{Mondal2015a}. 
These results thus beg again the question what the origin of the observed coherent MOKE response is.
In a very recent work, Hinschberger and Hervieux \cite{hervieux2015} investigated the non-linear (third order) dielectric response tensor and its effect on coherent magneto-optical measurements. They showed that the laser field modifies the magnetic state of the material by inducing a molecular mean field that is related to spin-orbit coupling between spins and optical photons. However, the explanations of the differences in differential Kerr rotation within the first 100 fs are still an open question given the fact that the pump and probe pulses interact with each other and also with the material \cite{Palfrey1985}. 
In this article we investigate theoretically the influence of the pump-probe relative polarization for the TR-MOKE experiments 
at very short time-delay 
$\tau$. Starting from a relativistic quantum description we show that the difference in the TR-MOKE spectra for different configurations (parallel or perpendicular polarizations) of pump and probe beams can be explained by the relativistic spin-photon coupling.      
 We furthermore  investigate the role that could be played in such kind of experiment by the so-called inverse Faraday effect (IFE) \cite{Berritta2016}, through which a magnetization could be induced by the laser light.

\section{Relativistic light-spin coupling Hamiltonian}
Electrons are relativistic particles and can consequently be described by the fundamental Dirac equation \cite{strange98}. Here we consider electrons in magnetic materials, for which the Dirac-Kohn-Sham equation is the adequate starting point. This entails that the electron is described relativistically as a single particle in an effective potential which results from all electrons and nuclei present. The presence of an external electromagnetic field (e.g., light) implies that the momentum of the electrons is changed by the minimal coupling, i.e., $\bm{p}\rightarrow (\bm{p}-e\bm{A})$, where $\bm{A}(\bm{r},t)$ is the 
vector potential and $e$ is the electronic charge, $e<0$. Applying an unitary, Foldy-Wouthuysen transformation \cite{foldy50,greiner00,Mondal2015a,hinschberger12} and keeping only up to the first order terms in $\frac{1}{c^2}$, we obtain the Hamiltonian for the large wavefunction components of the Dirac bi-spinor \cite{Mondal2015a,Mondal2015b} 
\begin{align}
	\mathcal{H}_{\rm FW} &=\frac{\left(\bm{p}-e\bm{A}\right)^{2}}{2m}+V + e\Phi -  \mu_{\rm B} \,\bm{\sigma}\cdot \bm{B} -\frac{\left(\bm{p}-e\bm{A}\right)^{4}}{8m^{3}c^{2}}-\frac{1}{8m^{2}c^{2}}\left(p^{2}V\right)-
\frac{e\hbar^{2}}{8m^{2}c^{2}}\bm{\nabla}\cdot\bm{E}\nonumber\\
& +\frac{i}{4m^{2}c^{2}}\bm{\sigma}\cdot\left(\bm{p}V\right)\times\left(\bm{p}-e\bm{A}\right)-\frac{e\hbar}{8m^{2}c^{2}}\bm{\sigma}\cdot\Big[ \bm{E}\times\left(\bm{p}-e\bm{A}\right)-\left(\bm{p}-e\bm{A}\right)\times\bm{E}\Big]\,.
\label{hamiltonian}
\end{align}
Here, $c$ is the speed of light, $m$ is the electron mass, $\mu_{\rm B}$ is the Bohr magneton, $\hbar$ is the reduced Plank constant and $\bm{\sigma}$ are the Pauli spin matrices. $V$ represents the crystal potential that is created by the nuclei and the other electrons and $\Phi$ defines the scalar potential of the externally applied electric field.
The involved fields are: $\bm{B}=\bm{\nabla}\times\bm{A}$ is the external magnetic field, $\bm{E}=-\frac{\partial \bm{A}}{\partial t} - \bm{\nabla}\Phi$ is electric field from the external source (e.g., light). We note that in the Hamiltonian we have left out the magnetic exchange field because it will not be relevant for the relativistic spin-light coupling of our interest; the full expression with exchange field can be found in Ref.\ \cite{Mondal2015a}. 
The second line of the above derived Hamiltonian has great importance for describing different relativistic spin-orbit-related effects. For example, it has been 
shown that the spin-orbit coupling terms that are linear in one of the \textit{external} fields explain the relativistic origin of the Gilbert damping tensor \cite{hickey09,Mondal2016}.
In a recent work we showed that the above Hamiltonian can be employed to investigate particular relativistic contributions to magneto-optical Kerr spectra \cite{Mondal2015a}.
Following the proposal of Bigot \textit{et al.} \cite{bigot09}, we computed the relativistic linear coupling of the laser's electromagnetic field to the electron's spin (and including the exchange field contributions), leading to spin-flip optical transitions, and found that these give only a very small contribution to the laser-induced change in the magneto-optical Kerr spectrum.
Recently, we have focused on the spin--electromagnetic field coupling terms that are quadratic in the external fields \cite{Mondal2015b}, which are the fields $\bm{E}\times\bm{A}$ and  $\bm{A}\times\bm{E}$ that appear in the last two terms of Eq.\ (\ref{hamiltonian}). As we showed, these last terms express the linear coupling of the optical angular momentum to the electron's spin \cite{Mondal2015b}, a property that had not been noticed in earlier work on relativistic quantum electrodynamics.
First, it is worth to notice that with the Foldy-Wouthuysen formalism we naturally recover these terms which ensure the gauge invariance of the Hamiltonian.
Second, the combined term that we refer to as Angular Magneto-Electric (AME) Hamiltonian \cite{Mondal2015b}, leads to several magneto-electric and magneto-optical phenomena (see, e.g. \cite{Raeliarijaona2013,Bellaiche2013,Walter2014}).
The electron spin dynamics induced by photons can be described qualitatively and quantitatively when this term is included along with the Pauli Hamiltonian especially for the strong fields' limit \cite{bauke14,BaukePRA2014}. In the following we investigate the influence of this relativistic term on femtosecond pump-probe magneto-optical spectroscopy.

Before doing so, let us mention that coherent ultrafast magnetism has extensively been studied recently along with the consideration of various relativistic spin-orbit terms \cite{Hervieux2016,hinschberger13,dixit13,Hinschberger2015}.  
Hinschberger {\it et al.}\ \cite{hervieux2015} considered the spin-orbit interaction terms between electromagnetic field due to charges and spins in the material and the fields from laser as: $\bm{\sigma} \cdot (\bm{E}_{\rm L} \times \bm{p})$, $\bm{\sigma} \cdot (\bm{E}_{\rm M} \times e\bm{A}_{\rm L})$ and $\bm{\sigma} \cdot (\bm{E}_{\rm L} \times e\bm{A}_{\rm M})$, where the subscripts `{L}' and `{M}' stand for the laser light and the material, respectively. These interactions, which can lead to light-induced spin flips during the excitation, have been taken into account, too, in a previous investigation of Vonesch and Bigot \cite{vonesch12}. A full numerical investigation of the influence of all relativistic terms that are \textit{linear} in the external fields on the magneto-optical response was also performed \cite{Mondal2015a}, but predicted only very small effects. However this does not solve the question: what is the role of pump and probe polarization during first 100 fs? As pointed out by Bigot {\it et al.} \cite{bigot09} the coherent magnetic signal appearing during pump-probe overlap strongly depends on the polarization state of both the pump and probe and a non-negligible difference is seen in the measurement within the first 100 fs (see Fig.\ \ref{fig1}). Thus, pump and probe interactions \cite{Palfrey1985} involving as well the spins are expected to play a role within these 100 fs because both pump and probe have a pulse width of 48 fs \cite{bigot09}. 
\section{Effect of relativistic light-spin coupling Hamiltonian in pump-probe experiments}
The AME Hamiltonian, which we intend to use to describe the femtosecond pump-probe experiments, appears \textit{quadratic} in the external electromagnetic  fields and can be rearranged as \cite{Mondal2015b,Paillard2016SPIE}
\begin{equation}
\mathcal{H} = \frac{e^2\hbar}{4m^{2}c^{2}}\bm{\sigma}\cdot\left(\bm{E}\times\bm{A}\right) .	
\label{hamil}
\end{equation}
Here we have used the Coulomb gauge, thus  $\bm{E} = -\frac{\partial \bm{A}}{\partial t }$ is the externally applied electric field and $\bm{A}(\bm{r},t)$ the external vector potential.
This Hamiltonian can naturally reproduce the relativistic contribution to the inverse Faraday effect \cite{Mondal2015b,Paillard2016SPIE} and other spintronic effects like the planar Hall effect \cite{Walter2014}, anomalous Hall effect \cite{Bellaiche2013} etc. It has also been shown that the AME Hamiltonian is at the origin of the optical spin-orbit torque, that is,  the spin torque exerted by the optical angular momentum on the electron's spin angular momentum \cite{Mondal2016}. In a recent study the AME Hamiltonian was considered in  the context of the Landau-Lifshitz-Gilbert spin dynamics with an additional optomagnetic field derived from the AME light-spin coupling 
 \cite{paillard-thesis}. 
Although several spin-orbit coupling terms were recently considered \cite{bigot09,vonesch12,Hervieux2016,hinschberger13,dixit13,Hinschberger2015} the Hamiltonian in Eq.\ (\ref{hamil}) was not yet treated. Following the idea of Bigot \textit{et al.} \cite{bigot09}, it contains the electric fields from femtosecond lasers, that could be from pump or probe pulses. 
Assuming a monochromatic plane electromagnetic wave propagating in the material one arrives at~\cite{Mondal2015b}:
\begin{equation}
	\mathcal{H} = \frac{e^2\hbar}{8m^{2}c^{2}\omega}\bm{\sigma}\cdot\mathcal{R}\left[-i\left(\bm{E}\times\bm{E}^\star\right)\right]\,,
	\label{hamiltonian1}
\end{equation}
where $\mathcal{R}$ defines the real part entering into the Hamiltonian.
During first few femtoseconds we treat the fields $\bm{E}= \bm{E}_{\rm pu}+ \bm{E}_{\rm pr}$ as a contribution from both pump and probe pulses and consequently their interaction.
Considering the pump and probe electric fields as plane waves with delayed Gaussian convolutions, we can write
\begin{align}
	\bm{E}_{\rm pu} & = \bm{E}^0_{\rm pu}\, e^{i\left(\bm{k}_{\rm pu}\cdot \bm{r} - \omega_{\rm pu} t\right)} \, e^{- \frac{t^2}{2\Gamma^2}} ,\nonumber\\ 
	\bm{E}_{\rm pr} & = \bm{E}^0_{\rm pr}\, e^{i\left(\bm{k}_{\rm pr}\cdot\bm{r} - \omega_{\rm pr} (t-\tau)\right)} \, e^{-\frac{ (t-\tau)^2}{2\gamma^2}} ,
	\label{e-fields}
\end{align} 
where, without loss of generality we can assume $\bm{E}_{\rm pu/pr}^0={E_{\rm pu/pr}^0}(\hat{\bm{e}}_x+e^{i\eta_{\rm pu/pr}}\hat{\bm{e}}_y)/ {\sqrt{2}}$ in which $\eta_{\rm pu/pr}$ define the ellipticity of the pump and probe beam, respectively.
Ordinarilly, in the experiment the pump intensity is always much higher than the probe intensity, $\vert \bm{E}^0_{\rm pu}\vert  >> \vert \bm{E}^0_{\rm pr}\vert $. The angular frequencies of both the pulses are denoted as $\omega_{\rm pu}$ and $\omega_{\rm pr}$. The parameters $\Gamma$ and $\gamma$ determine the pulse duration of the pump and probe, respectively, and $\tau$ accounts for the time delay between pump and probe.
In principle, the frequency, thus the energy of both the signals could be different e.g., the pump pulses has a central wavelength of 800 nm whereas the probe pulses can be generated by frequency doubling \cite{shen84}, i.e., second harmonic generation (SHG)  (see, e.g., \cite{Cinchetti2006,Roth2008JPD,Koopmans2010}). 
Here we assume the angular frequencies to be the same, $\omega_{\rm pu}=\omega_{\rm pr} = \omega$, 
 as in the measurements of Ref.\ \cite{bigot09}, 
which also implies $k_{\rm pu}=k_{\rm pr}$.
Lastly, the wave-vector $\bm{k}$ appearing in Eq.\ (\ref{e-fields}) is related to the refractive index $\bm{n}$, $\bm{k}=\frac{\omega \bm{n}}{c}$, and is hence a materials' dependent quantity.
Using these considerations the Hamiltonian (\ref{hamiltonian1}) can be simplified as,
\begin{align}
	\mathcal{H}&=\frac{e^2\hbar}{8m^{2}c^{2}\omega}\bm{\sigma}\cdot\mathcal{R}\left[-i\left(\bm{E}_{\rm pu}+\bm{E}_{\rm pr}\right) \times \left(\bm{E}_{\rm pu}^{\star}+\bm{E}_{\rm pr}^\star\right)\right]\nonumber\\
	& = \frac{e^2\hbar}{8m^{2}c^{2}\omega}e^{-\frac{ t^2}{2\Gamma^2}-\frac{ (t-\tau)^2}{2\gamma^2}}\bm{\sigma}\cdot\mathcal{R}\Big[-i(\bm{E}^{0}_{\rm pu}\times\bm{E}^{0\star}_{\rm pr}\,e^{-i\omega\tau}- \bm{E}^{0\star}_{\rm pu}\times
	\bm{E}^{0}_{\rm pr}\,e^{i\omega\tau}+\nonumber\\
&~~~~\bm{E}^{0}_{\rm pu}\times\bm{E}^{0\star}_{\rm pu}+\bm{E}^{0}_{\rm pr}\times\bm{E}^{0\star}_{\rm pr})\Big]\,,
		\label{energy}
\end{align}
The last two terms refer to the individual contributions of the pump and the probe pulses. These are different from zero only when these pulses have an elliptical polarization with non-zero ellipticity.
Following the re-writing of the AME Hamiltonian done in Ref.\ \cite{Mondal2015b}, the Hamiltonian adopts the form of an optical Zeeman field coupled with the spin of the electrons,
$-g \mu_{\rm B} \bm{\sigma} \cdot\bm{B}_{\rm opt} $, where the induced opto-magnetic field will be given by
\begin{align}
\bm{B}_{\rm opt}  =&\frac{e^2\hbar}{8m^{2}c^{2}\omega g\mu_{\rm B}}e^{-\frac{ t^2}{2\Gamma^2}-\frac{ (t-\tau)^2}{2\gamma^2}}\nonumber\\
&\times\mathcal{R}\Big[i(\bm{E}^{0}_{\rm pu}\times\bm{E}^{0\star}_{\rm pr}\,e^{-i\omega\tau}- \bm{E}^{0\star}_{\rm pu}\times\bm{E}^{0}_{\rm pr}\,e^{i\omega\tau})+\bm{E}^{0}_{\rm pu}\times\bm{E}^{0\star}_{\rm pu}+\bm{E}^{0}_{\rm pr}\times\bm{E}^{0\star}_{\rm pr})\Big]\,.
\end{align}
{On the other hand, considering $\omega_{\rm pu}\neq \omega_{\rm pr}$ \cite{Koopmans2010}, Eq.\ (\ref{hamiltonian1}) is no longer valid for bichromatic light and in this case these two quantities will be given as
\begin{align}
	\mathcal{H}= &\frac{e^2\hbar}{8m^{2}c^{2}}\bm{\sigma}\cdot\left\{\mathcal{R}[\bm{E}_{\rm pu}+\bm{E}_{\rm pr}]\times\left[-i\left(\frac{\bm{E}_{\rm pu}}{\omega_{\rm pu}}+\frac{\bm{E}_{\rm pr}}{\omega_{\rm pr}}\right)\right]\right\}\nonumber\\
=&-g\mu_B\sigma_{z}\Big(-\frac{e^2\hbar}{4m^{2}c^{2}g\mu_{\rm B}}\Big)\Big\{e^{-\frac{ t^2}{2\Gamma^2}-\frac{ (t-\tau)^2}{2\gamma^2}}\frac{E_{\rm pu}^0 E_{\rm pr}^0}{2}\nonumber\\
&\Big[\frac{1}{\omega_{\rm pr}}(\cos\phi_{\rm pu}\sin(\phi_{\rm pr}+\eta_{\rm pr})-\sin\phi_{\rm pr}\cos(\phi_{\rm pu}+\eta_{\rm pu}))+\nonumber\\
&\frac{1}{\omega_{\rm pr}}(\cos\phi_{\rm pr}\sin(\phi_{\rm pu}+\eta_{\rm pu})-\sin\phi_{\rm pu}\cos(\phi_{\rm pr}+\eta_{\rm pr}))\Big]-\nonumber\\
&-e^{- \frac{t^2}{\Gamma^2}}\frac{(E_{\rm pu}^0)^2}{\omega_{\rm pu}}\sin\eta_{\rm pu}-e^{- \frac{(t-\tau)^2}{\gamma^2}}\frac{(E_{\rm pr}^0)^2}{\omega_{\rm pr}}\sin\eta_{\rm pr}\Big\},
\end{align}	
	with the part after $\sigma_z$ being the now relevant opto-magnetic field $B_{\rm opt}$;
further, we abbreviated some phases as,
$\phi_{\rm pu} = \bm{k}_{\rm pu}\cdot\bm{r}-\omega_{\rm pu}t$ and 
$\phi_{\rm pr} = \bm{k}_{\rm pr}\cdot\bm{r}-\omega_{\rm pr}t-\omega_{\rm pr}\tau$.
The difference between the two angular frequencies is denoted as $\Delta\omega=\omega_{\rm pu}-\omega_{\rm pr}$.
In what follows we compare the Hamiltonian and the induced opto-magnetic field for two cases, a linearly or circularly polarized pump and a linearly polarized probe beam.     
 
\subsection{Pump and probe are both linearly polarized laser beams:}
First, we investigate the two cases described in the article by Bigot {\it et al.} \cite{bigot09}
\begin{enumerate}
	\item \underline{Pump beam is parallel to the probe beam ($ \bm{E}_{\rm pu}\parallel \bm{E}_{\rm pr} $), i.e.\ $\theta_{\rm PP} = 0^0$:}
As the pump and probe beams are parallel to each other and assuming that the beams are propagating in the $z$ direction, the simplest form one can have for linearly polarized pump and probe beams is:
\begin{align}
	\bm{E}^{0}_{\rm pu} = E_{\rm pu}^{0}\hat{\bm{e}}_{x},  \quad \bm{E}^{0}_{\rm pr} = E_{\rm pr}^{0}\hat{\bm{e}}_{x} \quad {\rm or} \quad \bm{E}^{0}_{\rm pu} = E_{\rm pu}^{0}\hat{\bm{e}}_{y}, \quad  \bm{E}^{0}_{\rm pr} = E_{\rm pr}^{0}\hat{\bm{e}}_{y}.
\end{align} 
For both sets of parallel polarization, the spin-photon interaction Hamiltonian does obviously not have any contribution because $\bm{E}^{0}_{\rm pu}\times \bm{E}^{0\star}_{\rm pr} = 0$ or $\bm{E}^{0\star}_{\rm pu}\times \bm{E}^{0}_{\rm pr} = 0$, giving $\mathcal{H}= 0$.
Consequently, there will not be any induced opto-magnetic field present in this case. 
 
\item \underline{Pump beam perpendicular to the probe beam ($ \bm{E}_{\rm pu}\perp \bm{E}_{\rm pr} $), i.e.\ $\theta_{\rm PP} = 90^0$:}

We consider again the electromagnetic wave propagating along the $z$-direction and electric fields that have components in $xy$ plane. For sake of simplicity, let us take the perpendicular pump and probe beams as:
\begin{align}
	\bm{E}^{0}_{\rm pu} = E_{\rm pu}^{0}\hat{\bm{e}}_{x}, \quad
	 \bm{E}^{0}_{\rm pr} = E_{\rm pr}^{0}\hat{\bm{e}}_{y} .
\end{align}		
Using  $\bm{E}^{0}_{\rm pu}\times\bm{E}^{0\star}_{\rm pr}=E_{\rm pu}^{0}E_{\rm pr}^{0\star}\,\hat{\bm{e}}_{z}$ and $\bm{E}^{0\star}_{\rm pu}\times\bm{E}^{0}_{\rm pr}=E_{\rm pu}^{0\star}E_{\rm pr}^{0}\,\hat{\bm{e}}_{z}$, the corresponding energy in terms of the Hamiltonian will be given as
\begin{align}
	\mathcal{H}&= \frac{e^2\hbar\vert E_{\rm pu}^{0}E_{\rm pr}^{0}\vert}{8m^{2}c^{2}\omega} e^{-\frac{ t^2}{2\Gamma^2}-\frac{ (t-\tau)^2}{2\gamma^2}}\sigma_z \,\,\mathcal{R}\Big[-i\left(\,e^{-i\omega\tau} - \,e^{i \omega\tau}\right)\Big]\nonumber\\
	&=-\frac{e^2\hbar\vert E_{\rm pu}^{0}E_{\rm pr}^{0}\vert}{4m^{2}c^{2}\omega} e^{-\frac{ t^2}{2\Gamma^2}-\frac{ (t-\tau)^2}{2\gamma^2}}\sigma_z \,\,\sin (\omega \tau) .
\end{align}
This implies that for linearly polarized beams, where the pump and probe polarizations are perpendicular, the relativistic spin-photon coupling Hamiltonian does not vanish.
This difference in the light-spin Hamiltonian for parallel and perpendicular pump-probe setups could explain the difference in the measurement in the first 100 femtoseconds. To substantiate this, we evaluate the size of the corresponding opto-magnetic field in typical experiments.

The induced opto-magnetic field is 
\begin{align}
	\bm{B}_{\rm opt}&=\frac{e^2\hbar\vert E_{\rm pu}^{0}E_{\rm pr}^{0}\vert}{4m^{2}c^{2}\omega  g\mu_{B}}e^{-\frac{ t^2}{2\Gamma^2}-\frac{ (t-\tau)^2}{2\gamma^2}}\sin (\omega \tau) \, \hat{\bm{e}}_{z} .
\end{align}
The maximum value of the induced opto-magnetic field will have the magnitude   
\begin{align}
	B_{\rm opt}&=\frac{e^2\hbar\vert E_{\rm pu}^{0}E_{\rm pr}^{0}\vert}{4m^{2}c^{2}\omega g\mu_{\rm B}} = \frac{e\hbar}{4mc^2}\frac{\vert E_{\rm pu}^{0}E_{\rm pr}^{0}\vert}{\hbar\omega } .
\end{align}
We evaluate numerically the magnitude of this opto-magnetic field by considering typical values of the appearing quantities. We adopt for the pump and probe electric field amplitudes as $\vert E ^{0}_{\rm pu}\vert  = 4\times 10^{8}$ V/m and $\vert E ^{0}_{\rm pr}\vert  =  10^{7} $ V/m, consistent with the experiment \cite{bigot09,hervieux2015}. These values correspond to pump and probe intensities of 21.2 GW/cm$^2$ and 13.28 MW/cm$^2$, respectively. Further, the 48 fs duration of pump and probe pulses leads to the value $24\times 10^{-15}$ s for the parameters $\gamma$ and  $\Gamma$ s \cite{bigot09}. The wavelength of the experiment was centered at 798 nm \cite{bigot09}, giving a photon energy of $\hbar \omega$ = $1.55$ eV. These considerations result in an induced opto-magnetic field of 0.83 $\mu$T. This field will act as a torque on the in-plane magnetization of the 7.5-nm Ni film during the overlap of the pump and probe beams. 
\end{enumerate}

\subsection{Elliptically polarized pump and linearly polarized probe beam:}
Next we examine the situation for an elliptically polarized pump pulse and a linearly polarized probe beam. This configuration was employed by Kimel \textit{et al.}\  \cite{Kimel2005} when studying light-induced spin oscillations in DyFeO$_3$.  
As before, we assume the pump pulse to be a general, elliptically polarized ray, 
$	\bm{E}^{0}_{\rm pu} = \left(\hat{\bm{e}}_{x}+e^{i\eta}\hat{\bm{e}}_{y}\right) {E_{\rm pu}^{0}}/ {\sqrt{2}}$,
where  the ellipticity 
$\eta = \pi/2$ for right-circularly polarized light and $-\pi/2$ for left-circularly polarized light.
The probe pulse is  taken as linearly polarized light, with polarisation in the xy-plane, and propagating parallel to the pump pulse,
\begin{align}
	\bm{E}_{\rm pr}^{0} = E_{\rm pr}^{0}\,\hat{\bm{e}}_{x}.
\end{align}	
The Hamiltonian in Eq.\ (\ref{energy}) can now be rewritten as
\begin{align}
	\mathcal{H}&= 
\frac{e^2\hbar}{4m^{2}c^{2}\omega}\sigma_z\Big(\frac{\vert E_{\rm pu}^{0}E_{\rm pr}^{0}\vert}{2}e^{-\frac{ t^2}{2\Gamma^2}-\frac{ (t-\tau)^2}{2\gamma^2}}(\sin (\omega \tau)-\sin(\omega\tau+\eta ))
+e^{-\frac{t^2}{\Gamma^2}}\frac{\sin\eta}{\omega}(E_{\rm pu}^0)^2\Big)\,.
\end{align}
This is the general Hamiltonian for an elliptically polarized pump beam. Note that there are now two contributions to the opto-magnetic field, a pump-pump and a pump-probe cross contribution. Considering a right-circularly polarized pump pulse, 
the Hamiltonian is
\begin{align}
	\mathcal{H}=\frac{e^2\hbar}{4m^{2}c^{2}\omega}\sigma_z\Big(-\frac{\vert E_{\rm pu}^{0}E_{\rm pr}^{0}\vert}{\sqrt{2}}e^{-\frac{ t^2}{2\Gamma^2}-\frac{ (t-\tau)^2}{2\gamma^2}}\sin (\omega \tau +\pi/4)+e^{-\frac{t^2}{\Gamma^2}}\frac{\sin\eta}{\omega}(E_{\rm pu}^0)^2\Big) ,
\end{align}  
whereas for a left-circularly polarised pump 
the Hamiltonian $-\mathcal{H}$ is obtained.
This implies that the induced opto-magnetic field will have opposite directions along the beam's propagation direction for right and left circular pump pulses. 
 The maximum magnitude of the pump-probe opto-magnetic field is in this configuration given by
\begin{align}
	B_{\rm opt}^{\rm pu-pr}=\frac{e^2\hbar\vert E_{\rm pu}^{0}E_{\rm pr}^{0}\vert}{4\sqrt{2}m^{2}c^{2}\omega g \mu_{\rm B}}.
\end{align}
Adopting again typical experimental parameters \cite{Kimel2005}, as pump and probe photons energy of $\hbar \omega = 1.55$ eV and 200 fs pulse duration, which gives $\gamma = \Gamma = 100\times 10^{-15}$ s.  The experimental intensity ratio of pump and probe beams of about 100 provides the amplitude of the pump electric field as $\vert E_{\rm pu}^{0}\vert =10 \times \vert E_{\rm pr}^{0}\vert$. The pump-probe opto-magnetic field is then given as a function of the pump intensity, $I_{\rm pu}$ as
\begin{align}
	B_{\rm opt}^{\rm pu-pr}=\frac{e^2\hbar I_{\rm pu}}{20\sqrt{2}m^2 c^{3}\epsilon_0\omega g\mu_{\rm B}},
\end{align} 
where the intensity is defined as
\begin{equation}
I_{\rm pu}=\frac{1}{2}\epsilon_0c\vert E_{\rm pu}^0\vert^2 .
\end{equation}
Considering similar intensities as for the laser of the previous experiment \cite{bigot09} we obtain a value for the opto-magnetic field of $B_{\rm opt}^{\rm pu-pr}=0.58$ $\mu$T.
On the other hand, the effect of the pump-pump cross term is given by
\begin{equation}
B_{\rm opt}^{\rm pu-pu}=\frac{e^2\hbar\vert E_{\rm pu}^0\vert^2}{4m^2c^2\omega g\mu_{\rm B}} =
\frac{ e^2 \hbar I_{\rm pu}}{2 m^2 c^3 \epsilon_0 \omega g\mu_{\rm B}} ,
\end{equation}
which leads to a much larger contribution to the opto-magnetic field, $B_{\rm opt}^{\rm pu-pu}=33.2$ $\mu$T.
In the experiment described in Ref. \cite{Kimel2005} the authors used a laser with a duration $\Gamma=200$ fs and a fluence $F=500$ mJ/cm$^2$. Starting from this values we can calculate the intensity of the laser as:
\begin{equation}
I_{\rm pu}=\frac{F}{\Gamma}
\end{equation}
which provides us with a value for the laser intensity $I=2.5$ TW/cm$^2$ corresponding to an opto-magnetic field due to the pump $B_{\rm opt}^{\rm pu-pu}=3.9$ mT and, assuming a probe beam with an intensity 100 times smaller, an opto-magnetic field due to the interaction between pump and probe $B_{\rm opt}^{\rm pu-pr}=0.28$ mT.

\section{Role played by the inverse Faraday effect}

An important role in the interaction between pump and probe pulses, in analogy with the relativistic effect just described, is played by the inverse Faraday effect (IFE), which was expected first for a circularly polarized pump pulse \cite{Kimel2005}.
The IFE is the strong-field opto-magnetic effect in which coherent laser light induces a magnetization in a material (for discussions of the IFE, see e.g., Refs.\ \cite{woodford09,popova12,taguchi11b,barbalinardo14,Freimuth2016,Berritta2016}). 
The IFE can be described as a non-linear, quadratic response to the coupling term $\bm{p} \cdot \bm{A}$ appearing in the Hamiltonian (\ref{hamiltonian}). This is obviously a different coupling term from the  above discussed AME spin-photon term $\bm{\sigma} \cdot (\bm{E} \times \bm{A})$. Nonetheless, both terms lead to an induced magnetization or opto-magnetic field proportional to $I_{\rm pu}$; in the case of the IFE the spin-orbit interaction is not in the operator $\bm{p} \cdot \bm{A}$ but in the electron wavefunctions, whereas for the AME the spin-photon interaction itself is a relativistic operator.

Considering again the two configurations studied by Bigot \textit{et al.}\ \cite{bigot09}, it would at first sight appear that the employed linearly polarized pump pulses cannot induce any magnetization, in contrast to circularly polarized pulses. However, in disparity to this expectation, recent \textit{ab initio} calculations have shown that a magnetization can be indeed induced in ferromagnets as well with linearly polarized light \cite{Berritta2016}. 
In the discussed TR-MOKE experiment the linearly polarized pump would for both configurations $\bm{E}_{\rm pu}\perp\bm{E}_{\rm pr}$ and $\bm{E}_{\rm pu}\parallel \bm{E}_{\rm pr}$ induce a same magnetization along the pump's wave-vector. Since this imparted magnetization is probed in polar MOKE geometry, a difference between two measurements with linear probe polarizations that differ by $90^{\circ}$ is not expected. Consequently, this suggests the relativistic spin-photon coupling to be at the origin of the TR-MOKE response difference observed in the first 100 fs. 

The situation is different for the second analyzed experiment, in which the pump is circularly polarized \cite{Kimel2005}. In this case both the IFE and the AME spin-photon term could lead to an induced magnetization or opto-magnetic field. As shown recently the IFE depends on optical transitions and is not an absorption-free quantity \cite{Berritta2016}. In a wide band-gap material which is pumped with a laser energy substantial below the band gap the IFE magnetization is likely very small and an induced opto-magnetic field in the low absorption region would hence stem from the AME spin-photon coupling only.

\section{Conclusions}
To understand the origin of the coherent ultrafast magnetization response observed in femtosecond pump-probe magneto-optical measurements \cite{bigot09} we have derived a relativistic Hamiltonian which accounts for the relativistic interaction of the spins of the electrons in a material with the electromagnetic fields of the beams. This Hamiltonian naturally leads to a combined interaction of both the pump and probe beams during their overlap in the time domain in a pump-probe experiment. For linearly polarized pump and probe beams, there is a induced opto-magnetic field when the pump and probe polarizations are perpendicular to each other, but not when they  are parallel to one another. The value of the induced opto-magnetic field is estimated to be of the order of a few $\mu$T for the employed fluencies. 
This suggest that the relativistic light-spin interaction can provide an explanation of the observed coherent ultrafast magnetic response, as idea which was originally conjectured by Bigot \textit{et al.} \cite{bigot09}. Our investigation however identifies the relativistic term that is quadratic in the external applied fields as the responsible one, whereas other recent investigations considered coupling terms that are linear in one of the external fields \cite{vonesch12,Hervieux2016,hinschberger13,dixit13,Hinschberger2015,Mondal2015a}.
We further investigated the configuration where the pump beam is elliptically polarized and the probe beam is linearly polarized. 
We have shown that in this case the there exists an induced opto-magnetic field that follows a linear relationship with the pump fluence. This finding is consistent with magneto-optical measurements of the amplitude of spin oscillations which were found to increase linearly with the pump fluence \cite{Kimel2005}. We estimate that, for the employed pump fluences, the AME spin-photon coupling could induce a sizable opto-magnetic field of up to 3.9 mT. In the pump-probe configuration with a circularly polarized pump there can be additional contributions due to the inverse Faraday effect, which would depend on the amount of absorption in the material.
\section{Acknowledgments}
 We thank P.\ Maldonado, A.\ Aperis and A.\ K.\ Nandy for valuable discussions.
 This work has been supported by 
the Swedish Research Council (VR), the Knut and Alice Wallenberg Foundation (Grant No.\ 2015.0060), and the Swedish National  Infrastructure for Computing (SNIC).
\bibliographystyle{iopart-num}
\providecommand{\newblock}{}

\end{document}